\documentclass[10pt]{iopart}
\usepackage[dvips]{graphicx}
\usepackage{iopams}

\def\calh{{\mathcal H}} 
 
\newcommand{\He}{\mbox{\rm He}}

\begin{document} 

\title{Distribution of bipartite entanglement for random pure states}
\author{Olivier Giraud} 
\address{Laboratoire de Physique Th\'eorique, UMR 5152 du CNRS, 
Universit\'e Paul Sabatier, 31062 Toulouse Cedex 4, France}
\ead{giraud@irsamc.ups-tlse.fr}
\date{\today} 
\begin{abstract}
We calculate analytic expressions for the distribution of
bipartite entanglement for pure random quantum states. All moments
of the purity distribution
are derived and an asymptotic expansion for the distribution
itself is deduced. An approximate expression for moments
and distribution of Meyer-Wallach entanglement for random pure states
is then obtained.
\end{abstract}
\maketitle

%%%%%%%%%%%%%%%%%%%%%%%%%%%%%%%%%%%%%%%%%%%%%%%%%%%%%%%%%%%%%%%%%%%%% 

%**************************************************************
\section*{Introduction}
%**************************************************************
The question of generating and 
measuring entanglement in multipartite quantum systems has 
become of greater interest with the development of the field of
quantum information. Entanglement generation is an important aspect of 
several quantum information processes, such as superdense coding 
\cite{HarHayLeu}, quantum communication \cite{HayLeuShoWin},
or quantum data hiding \cite{DivLeuTer}.
Various methods have been proposed in order to generate highly entangled 
quantum states, based on pseudo-random unitary operators \cite{WeiHel}
or on the entangling power of chaotic quantum maps \cite{BanLak, ScoCav} or
intermediate quantum maps \cite{GirGeo}.
Entanglement generation by means of pseudo-random unitary operators 
or chaotic quantum maps relies upon the fact that unitary evolution 
of any initial state leads to states whose entanglement properties
are close to those of random states, in particular to highly entangled
 states.\\
In order to quantify the entanglement of a state, or the entangling 
power of an operator,
a number of entanglement measures 
have been proposed, based either on quantum information theory or on 
thermodynamical considerations:
entanglement of formation and distillable entanglement
\cite{BenDivSmoWoo}, relative entropy 
\cite{VedPleRipKni,VedPleJacKni,VedPle}, $n$-tangle \cite{WonChr},
concurrence \cite{HilWoo}. 
For bipartite entanglement of pure states, these measures
all reduce to the entropy of entanglement \cite{BenBerPopSch}, 
which can be proved to be a unique entanglement measure 
\cite{PopRoh, DonHorRud}. 
The entropy of entanglement corresponds to the von Neumann entropy
of the partial density matrix obtained by tracing over one subsystem.
 Rather than the von Neumann entropy itself,
one often prefers to consider the purity $R$, which corresponds 
(up to constants) to 
the so-called linear entropy, that is the 
first-order term in the expansion of the von Neumann entropy 
around its maximum. To quantify the degree of entanglement of 
multipartite pure states, one measure commonly used, based on purity,
 is the measure 
proposed by Meyer and Wallach in \cite {MeyWal}. It consists in
taking the average of the bipartite entanglement of one qubit with 
all others, measured by the purity (see Equation (\ref{mwent})) \cite{Bre}. 
Meyer-Wallach entanglement was used e.g. to quantify entanglement
generation for pseudo-random operators \cite{WeiHel}
or intermediate or chaotic quantum maps \cite{GirGeo, Sco, ZanZalFao}.\\
The study of purity or Meyer-Wallach entanglement
 is of particular interest for random
quantum states. Random pure states as column vectors of
random unitary matrices distributed according to the invariant Haar measure 
can be shown to be entangled with high probability.
Various analytical calculations have been carried out to characterize 
entanglement properties of random states. Expressions for the 
first moment of the purity 
have been obtained by Lubkin \cite{Lub}; the second and third 
moments have been derived in \cite{ScoCav}, following earlier work 
\cite{Sen}. The average entropy has been obtained in \cite{Pag}.
Statistical properties of entanglement measures
for random density matrices were obtained in \cite{SomZyc}-\cite{CapSomZyc}.
The average value for each Schmidt coefficient of a random pure state
has been calculated in \cite{Zni}.
In \cite{MalMenLen}, the average entropy of a subsystem was obtained from 
the average Tsallis entropy \cite{Tsa}.\\
To further characterize entanglement 
of random pure states, our aim here is to give an exact 
expression for all moments
of the probability density distribution $P(R)$ of the purity for a bipartite
random pure state. 
Since the probability distribution $P(R)$ is defined over a bounded 
interval ($R$ is bounded), 
the knowledge of all moments determines uniquely the probability
distribution \cite{Hau}. There are various techniques to obtain a 
function approximating the exact probability density distribution 
in a controlled way (that is, by an expansion
where the error can be bounded) from the knowledge of its
moments. In \cite{Gre} an algorithm was given
to construct polynomials converging to the probability distribution.
We will rather follow \cite{BliMoe}, where the asymptotic expansion 
for nearly gaussian distributions is calculated at all orders. 
In Section \ref{bipartite}
the moments $\langle R^n \rangle$ for the distribution  $P(R)$ are
calculated, and the construction of the asymptotic expansion of  $P(R)$
at all orders from its moments is recalled.
In Section \ref{multipartite} the approximate moments 
$\langle Q^n \rangle$ for the distribution  $P(Q)$ 
are derived. For both distributions, the
 moments are expressed as sums involving 
a finite number of combinatorial terms and can be 
easily calculated effectively. As an illustration, we give 
the first values of the cumulants and calculate the
probability density distribution expansion for Meyer-Wallach entanglement.

%**************************************************************
\section{Bipartite entanglement for random pure states}
\label{bipartite}
%**************************************************************
Let $\Psi$ be a pure state belonging to a Hilbert space 
$\calh_A\otimes\calh_B$, where $\calh_A$ and $\calh_B$ are 
spanned respectively by $\{|a_i\rangle\}_{1\leq i\leq p}$ and 
$\{|b_i\rangle\}_{1\leq i\leq q}$. We assume that $p\leq q$.
 Let $x_i$ be the Schmidt coefficients
for $\Psi$. That is, the state $\Psi$ has a Schmidt decomposition
(see e.g. \cite{NieChu})
\begin{equation}
|\Psi\rangle=\sum_{i=1}^{p}\sqrt{x_i}|a_i\rangle\otimes|b_i\rangle.
\end{equation}
The bipartite entanglement measure for $\Psi$ can be expressed 
throught Schmidt coefficients $x_i$. 
The entropy of entanglement is the Shannon entropy of the $x_i$'s:
$S(\Psi)=-\sum_{i=1}^{p}x_i \ln x_i$. The
  purity $R(\Psi)$ of the state $\Psi$ can be expressed as
\begin{equation}
\label{purity}
R(\Psi)=\sum_{i=1}^{p}x_i^2.
\end{equation}
For random states the Schmidt coefficients are distributed according
to the density
\begin{equation}
\label{density}
P(x_1, \ldots, x_p)=\mathcal{N}\prod_{1\leq i<j\leq p}(x_i-x_j)^2
\prod_{1\leq k\leq p}x_k^{q-p}\;\delta\left(1-\sum_{i=1}^{p}x_i\right)
\end{equation}
for $x_i\in[0,1]$, with some normalisation factor $\mathcal{N}$
 \cite{LloPag, ScoCav}.
The $n$-th moment of the purity is then given by
\begin{eqnarray}
\label{defrn}
\langle R^n \rangle=
\mathcal{N}\int_0^1 dx_1\ldots dx_p\prod_{1\leq i<j\leq p}(x_i-x_j)^2
\prod_{1\leq k\leq p}x_k^{q-p}\\
\nonumber\hspace{3cm}\times
\left(x_1^2+x_2^2+\cdots+x_p^2\right)^n\delta\left(1-\sum_{i=1}^{p}x_i\right).
\end{eqnarray}
The calculation of $\langle R^n \rangle$ requires the evaluation of integrals
of the form
\begin{equation}
\label{inti}
I({\bf n})=\int_0^1 dx_1\ldots dx_p V({\bf x})^2 x_1^r\ldots x_p^r
\delta\left(1-\sum_{i=1}^{p}x_i\right)f_{{\bf n}}({\bf x}),
\end{equation}
where $r=q-p$, ${\bf x}=(x_1, \ldots, x_p)$ and ${\bf n}=(n_1, \ldots, n_p)$.
The function $f_{\bf n}({\bf x})=\{x_1^{n_1}x_2^{n_2}\ldots x_p^{n_p}$
+ all permutations of the $n_i\}$ is a symmetric function of the $x_i$, 
and $V$ is the Vandermonde determinant
\begin{equation}
\label{vandermonde}
V({\bf x})=\prod_{1\leq i<j\leq p}(x_i-x_j).
\end{equation}
The integral $I({\bf n})$ is evaluated in the appendix and yields
\begin{equation}
\label{in}
I({\bf n})=\frac{p!\prod_{i=1}^{p}(r+n_i+i-1)!}{(p^2+r p+\sum_i n_i-1)!}
\prod_{i<j}(n_j-n_i+j-i)+\textrm{perm.},
\end{equation}
where ''$+$perm'' indicates that the expression \eref{in} is a sum over
all permutations of the $n_i$.
Now the function $f$ for a given $\langle R^n \rangle$ is obtained
by multinomial expansion of the term 
\begin{equation}
(x_1^2+\cdots+x_p^2)^n=\sum_{n_1+n_2+\cdots+n_p=n}
\frac{n!}{n_1!n_2!\ldots n_p!}x_1^{2n_1}x_2^{2n_2}\ldots x_p^{2n_p}.
\end{equation}
The normalization constant  $\mathcal{N}$ in \eref{defrn}
 is given by the choice ${\bf n}={\bf 0}$ in \eref{inti},
i.e. $\mathcal{N}=1/I({\bf 0})$. This leads to
\begin{eqnarray}
\langle R^n \rangle=\frac{(p^2+r p-1)!}{(p^2+r p+2n-1)!}
\sum_{n_1+n_2+\cdots+n_p=n}\frac{n!}{n_1!n_2!\ldots n_p!}\nonumber\\
\hspace{1.3cm}\times
\frac{\prod_{i=1}^{p}(r+2n_i+i-1)!}{\prod_{i=1}^{p}(r+i-1)!}
\prod_{1\leq i<j\leq p}\frac{2n_j-2n_i+j-i}{j-i}.
\end{eqnarray}
Replacing $r$ by its value $q-p$ and correspondingly changing all
indices $i$ to $p+1-i$ (and $n_i$ to $n_{p+1-i}$), one 
finally obtains
\begin{eqnarray}
\label{rnferme}
\langle R^n \rangle=\frac{(p q-1)!}{(p q+2n-1)!}
\sum_{n_1+n_2+\cdots+n_p=n}\frac{n!}{n_1!n_2!\ldots n_p!}\nonumber\\
\hspace{2cm}\times
\prod_{i=1}^{p}\frac{(q+2n_i-i)!}{(q-i)!i!}
\prod_{1\leq i<j\leq p}(2n_i-i-2n_j+j).
\end{eqnarray}
Note that one can cast \eref{rnferme} into an expression more symmetric 
in $p$ and $q$ by noting that
\begin{equation}
\frac{\prod_{i<j}(2n_i-i-2n_j+j)}{\prod_{i=1}^{p}(p+2n_i-i)!}=
\prod_{j=1}^{p}\left[\frac{1}{(2n_j)!}
\prod_{i=1}^{j-1}\left(1-\frac{2n_j}{2n_i+j-i}\right)\right],
\end{equation}
yielding
\begin{eqnarray}
\label{rnsymmetric}
\langle R^n \rangle=\frac{(p q-1)!}{(p q+2n-1)!}
\sum_{n_1+n_2+\cdots+n_p=n}\frac{n!}{n_1!n_2!\ldots n_p!}\\
\hspace{1cm}\times\prod_{n_i\neq 0}
\left[\frac{(q+2n_i-i)!(p+2n_i-i)!}{(q-i)!(p-i)!(2n_i)!}
\prod_{j=1}^{i-1}\left(1-\frac{2n_j}{2n_i+j-i}\right)\right].\nonumber
\end{eqnarray}
Equation \eref{rnferme} is a closed expression, 
involving only a finite sum over partitions of $n$
into numbers greater or equal to 0. Note that the order of the $n_i$ 
matters: for instance
for $p=2$ and $n=2$ the sum will involve three terms
$(n_1, n_2)=(2,0)$, $(1,1)$ and $(0,2)$. 
These partitions can be easily generated for any $n$ by some suitable
algorithm (see e.g. \cite{BliMoe} for such an algorithm generating 
the partitions required).
From Equation \eref{rnferme} one can get the expressions
for the cumulants of the distribution $P(R)$. Indeed, 
given the moments $\mu_n$ of a distribution the $n$-th cumulant
$\kappa_n$ reads (see e.g. \cite{BliMoe})
\begin{equation}
\label{moment-cumulant}
\kappa_n=n!\sum_{\{k_m\}}(-1)^{r-1}(r-1)!\prod_{m=1}^n\frac{1}{k_m!}
\left(\frac{\mu_m}{m!}\right)^{k_m},
\end{equation}
where $r=k_1+\cdots+k_n$, and 
the sum runs over all $k_i\geq 0, 1\leq i\leq n$ such that
$k_1+2k_2+...+nk_n=n$.
As an example, the first five cumulants read 
\begin{eqnarray}
\kappa_1 &=& \frac{p + q}{1 + p q}\\
\kappa_2 &=& \frac{2 (p^2-1)(q^2-1)}{(1 + p q)^2 
(2 + p q)(3 + p q)}\nonumber\\
\kappa_3 &=& \frac{8 (p^2-1)(q^2-1)(p + q)(-5+ p q)}
{(1 + p q)^3 (2 + p q)(3 + p q)(4 + p q)(5 + p q)}\nonumber\\
\kappa_4 &=& \frac{48 (p^2-1)(q^2-1)( p q-3)A_{p,q}}
{(1 + p q)^3 (2 + p q) (3 + p q) \prod_{i=1}^{7}(i + p q)}
\nonumber\\
\kappa_5 &=& \frac{384 (p^2-1)(q^2-1)(p + q)B_{p,q}}
{(1 + p q)^4 (2 + p q) (3 + p q)\prod_{i=1}^{9} (i + p q)}\nonumber
\end{eqnarray}
where $A_{p,q}$ and $B_{p,q}$ are polynomials in $p$ and $q$ defined by $A_{p,q}=28 - 112 p^2 - 153 p q -  79 p^3 q - 112 q^2 - 98 p^2 q^2- 11 p^4 q^2 - 79 p q^3 - 3 p^3 q^3 + p^5 q^3 - 11 p^2 q^4 +  4 p^4 q^4 
+ p^3 q^5$ and $B_{p,q}=3528 - 6552 p^2 - 6343 p q - 449 p^3 q - 6552 q^2 + 1545 p^2 q^2 + 1237 p^4 q^2- 449 p q^3 + 1164 p^3 q^3 + 132 p^5 q^3 + 1237 p^2 q^4 - 274 p^4 q^4- 41 p^6 q^4 + 132 p^3 q^5 - 93 p^5 q^5 + p^7 q^5 - 41 p^4 q^6 
+ 9 p^6 q^6 + p^5 q^7$.\\
As expected, $\kappa_1$ corresponds to Lubkin's expression \cite{Lub}
for the average purity.
For $n= 2, 3$ one recovers the expressions derived in \cite{ScoCav}.
For larger $n$ it is easy to generate the exact value for each cumulant.\\
In the case $p=2$, the analytic expression for 
the probability distribution $P(R)\;dR$ can 
easily be obtained analytically directly from \eref{purity}-\eref{density}.
It reads
\begin{equation}
\label{P2}
P(R)\;dR=A(1-R)^{q-2}\sqrt{2R-1}\;dR
\end{equation}
for $1/2\leq R\leq 1$, 0 otherwise ($A$ is the normalization factor). 
For $p\geq 3$, the asymptotic expansion of the distribution can be 
obtained (see \cite{BliMoe} and references therein) by 
Edgeworth expansion as a function of the normal distribution 
$Z(x)=\exp(-x^2/2)/\sqrt{2\pi}$,
the mean $\mu=\kappa_1$, the variance $\sigma^2=\mu_2-\mu_1^2=\kappa_2$, 
and rescaled cumulants $\gamma_r=\kappa_r/\sigma^{2r-2}$:
\begin{eqnarray}
\label{allorderexpansion}
P(R)=\frac{1}{\sigma}Z\left(\frac{R-\mu}{\sigma}\right)
\left[1+\right.\\
\nonumber
\hspace{2cm}\left.\sum_{s=1}^{\infty}\sigma^s
\sum_{\{k_m\}}\He_{s+2t}\left(\frac{R-\mu}{\sigma}\right)\prod_{m=1}^{s}\frac{1}{k_m!}
\left(\frac{\gamma_{m+2}}{(m+2)!}\right)^{k_m}\right].
\end{eqnarray}
For each $s$ the sum runs over $k_j\geq 0$ such that $\sum_j j k_j=s$,
and $t$ is defined by $t=\sum_j k_j$.
The $\He_n(x)$ are Chebyshev-Hermite polynomials defined by 
$\He_n(x)=(-1)^n e^{x^2/2}\partial^n e^{-x^2/2}$ (here $\partial$ is the
differential operator with respect to $x$) and correspond to
rescaled Hermite polynomials:
\begin{equation}
\label{hermite}
\He_n(x)=n!\sum_{k=0}^{[n/2]}\frac{(-1)^k x^{n-2k}}{k!(n-2k)!2^k}.
\end{equation}
Equations \eref{moment-cumulant}-\eref{hermite} together with
the knowledge of the moments \eref{rnferme}
allow to obtain explicitely the asymptotic expansion
of the probability density distribution at any order.

%**************************************************************
\section{Multipartite entanglement}
\label{multipartite}
\begin{figure}[ht]
\begin{center} 
\includegraphics*[width=.85\linewidth]{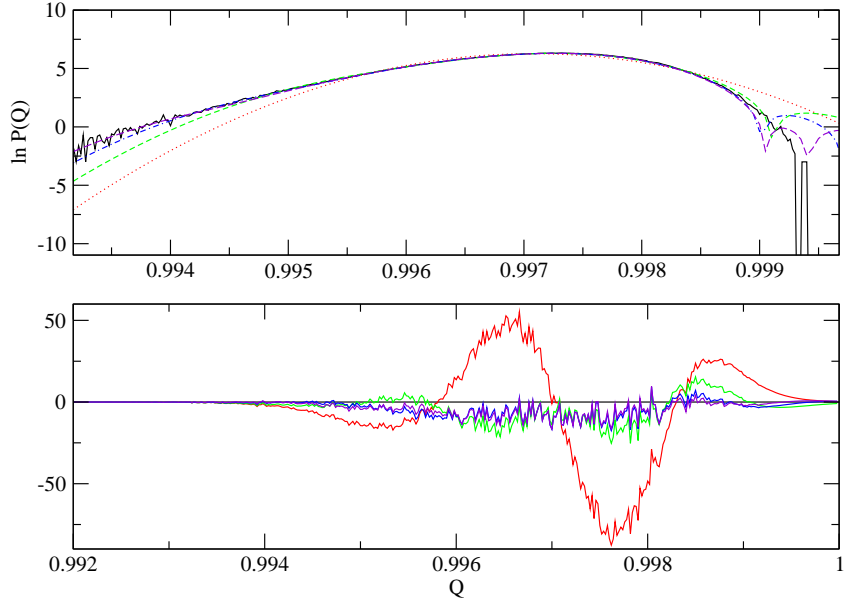}
\end{center} 
\caption{Probability density function $P(Q)$ of Meyer-Wallach entanglement $Q$ 
for random vectors of size $2^m$ for $m=10$. 
Top: $P(Q)$ in logarithmic scale. Bottom: differences between $P_s(Q)$ 
(analytical expansion at order $s$) and the numerical curve.
From bottom to top on the left axis of the top figure: truncation of 
the expansion at order 0 
(gaussian, red, dotted); order 1 (green, short dashed); 
order 2 (blue, dot-dashed); order 3 (purple, long dashed); numerical 
curve from column vectors of $1000$ random unitary matrices obtained
by Hurwitz parametrization (black, solid).}
\label{fig1}
\end{figure}

\begin{figure}[ht]
\begin{center} 
\includegraphics*[width=.85\linewidth]{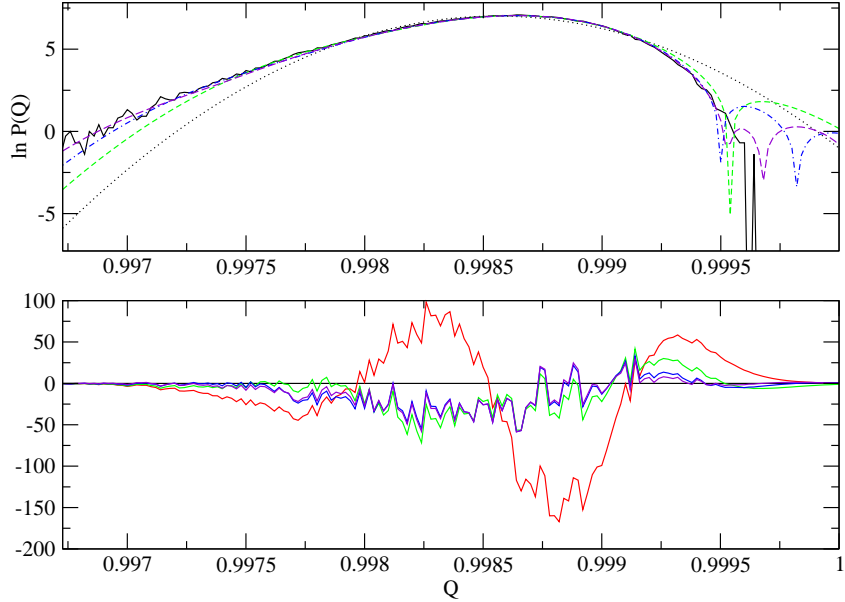}
\end{center} 
\caption{Same as Figure \ref{fig1} for $m=11$. Numerical curve averaged 
over 100 random unitary matrices.}
\label{fig2}
\end{figure}

%**************************************************************
The Meyer-Wallach entanglement of a pure
$M$-dimensional state $\Psi$ coded on $m$ qubits (with $M=2^{m}$)
 can be defined by
\begin{equation}
\label{mwent}
Q(\Psi)=2\left(1-\frac{1}{m}\sum_{i=1}^{m}R_k\right),
\end{equation}
where $R_k$ is the purity \eref{purity} of the $k$-th qubit \cite{Bre}.
In order to calculate $Q(\Psi)$ for bipartite
random pure states we need to calculate the average purity of a 
bipartite system belonging to a Hilbert space $\calh_A\otimes\calh_B$, where 
$\calh_A$ has dimension $p=2$ and $\calh_B$ has dimension $q=2^{m-1}=M/2$.
The moments $\langle R^n \rangle$ can be obtained in this case
either from Equation \eref{rnferme} or directly from the distribution
\eref{P2}. In both cases it leads to  
\begin{equation}
\label{rnp2}
\langle R^n \rangle=\frac{\Gamma(q+\frac{1}{2})}{\sqrt{\pi}2^{n-1}}
\sum_{k=0}^{n}{n \choose k}
\frac{\left(k+\frac{1}{2}\right)!}{\left(q+k-\frac{1}{2}\right)!}.
\end{equation}
The calculation of the moments $\langle Q^n \rangle$ involves 
terms of the form $\langle \left(\sum_i R_i\right)^k\rangle$.
These terms depend on correlations between the purities $R_i$.
However if we make the assumption that for two different qubits $i\neq j$ 
we have $\langle R_i R_j\rangle=\langle R_i\rangle\langle R_j\rangle$,
we get a distribution $P(Q)$ which turns out to be very close
to the numerical distribution obtained by generating random matrices. 
Making this assumption we get
\begin{equation}
\label{multiRn}
\langle \left(\sum_{i=1}^{m} R_i\right)^k\rangle=\hspace{-.5cm}
\sum_{k_1+k_2+\cdots+k_{m}=k}\frac{k!}{k_1!k_2!\ldots k_{m}!}
\langle R^{k_1} \rangle\langle R^{k_2}\rangle\ldots\langle R^{k_{m}}\rangle,
\end{equation}
The $n$-th moment is then
\begin{equation}
\label{Qnavant}
\langle Q^n \rangle=2^n
\sum_{k=0}^{n}{n \choose k}\frac{(-1)^k k!}{m^k}\hspace{-0.5cm}
\sum_{k_1+k_2+\cdots+k_{m}=k}\frac{\langle R^{k_1} \rangle}{k_1!}
\frac{\langle R^{k_2}\rangle}{k_2!}\ldots
\frac{\langle R^{k_{m}}\rangle}{k_m!}.
\end{equation}
Gathering together terms having the same exponents, we finally get
\begin{equation}
\label{Qn}
\langle Q^n \rangle=2^n
\sum_{k=0}^{n}{n \choose k}
\frac{(-1)^k k!}{m^k}
\sum_{\{r_k\}}\frac{m!}{r_1!r_2!\ldots r_{k}!(m-r)!}
\prod_{i=1}^{k}\left(\frac{\langle R^{i}\rangle}{i!}\right)^{r_i},
\end{equation}
where $r\equiv\sum r_i$ and $\langle R^n \rangle$ is given by \eref{rnp2}.
The sum runs over all $r_i\geq 0$ such that $\sum_j j r_j=k$.
From Equations \eref{rnp2} and \eref{Qn} one can now
obtain the cumulants for the distribution $P(Q)$ for an $m$-qubit
system ($M=2^m$). The first ones read
\begin{eqnarray}
\label{cumulantsQ}
\kappa^Q_1&=&\frac{M-2}{M+1}\\
\kappa^Q_2&=&\frac{6(M-2)}{(M+1)^2 (M+3)m}\nonumber\\
\kappa^Q_3 &=&\frac{24 (-M^2+7M-10)}
{(M+1)^3 (M+3)(M+5)m^2}\nonumber\\
\kappa^Q_4 &=&\frac{144(M^4-12M^3+6M^2+133M-210)}
{(M+1)^4(M+3)^2(M+5)(M+7)m^3}\nonumber\\
\kappa^Q_5 &=&-\frac{1152(1890-1763 M+337M^2 +78M^3 - 23M^4 +M^5)}
{(M+1)^5(M+3)^2(M+5)(M+7)(M+9)m^4}.\nonumber
\end{eqnarray}
We can use these approximate cumulants to obtain an analytical formula for 
$P(Q)$. Calculating the first terms in the asymptotic expansion 
\eref{allorderexpansion} we obtain (the first terms can be found
in \cite{AbrSte})
\begin{eqnarray}
\label{approx}
\nonumber
P(Q)&\sim& \frac{1}{\sigma}Z\left(\frac{Q-\mu}{\sigma}\right)
\left\{1+\frac{\tau_3}{6}\He_3\left(\frac{Q-\mu}{\sigma}\right)
\right.\\
\nonumber
&+&
\left[\frac{\tau_4}{24}\He_4\left(\frac{Q-\mu}{\sigma}\right)
+\frac{\tau_3^2}{72}\He_6\left(\frac{Q-\mu}{\sigma}\right)\right]\\
&+&\left[\frac{\tau_5}{5!}\He_5\left(\frac{Q-\mu}{\sigma}
\right)
+\frac{\tau_3\tau_4}{144}\He_7\left(\frac{Q-\mu}{\sigma}\right)
\right.
\\
&+&\left.\left.\frac{\tau_3^3}{1296}\He_9\left(
\frac{Q-\mu}{\sigma}\right)\right]+...\right\}
%&+&\left[\frac{\kappa_6}{6!\sigma^6}\He_6\left(\frac{Q-\mu}{\sigma}\right)
%+\left(\frac{\kappa_4^2}{1152\sigma^8}
%+\frac{\kappa_3\kappa_5}{720\sigma^8}\right)
%\He_8\left(\frac{Q-\mu}{\sigma}\right)\right.\nonumber\\
%&+&\left.\left.\left.\frac{\kappa_3^2\kappa_4}{1728\sigma^{10}}
%\He_{10}\left(\frac{Q-\mu}{\sigma}\right)+\frac{\kappa_3^4}{31104\sigma^{12}}
%\He_{12}\left(\frac{Q-\mu}{\sigma}\right)\right)\right]+...\right\},
\end{eqnarray}
with $\mu=\kappa^Q_1$, $\sigma=\sqrt{\kappa^Q_2}$ and
$\tau_i=\kappa_i/\sigma^i$, $i\geq 3$.
Figure \ref{fig1} displays the probability density function $P(Q)$ for
$m=10$ qubits as obtained by averaging over numerically generated random
matrices, together with the plot of analytical expression \eref{approx}
truncated at order 0 (gaussian), 1 (first line of \eref{approx}), 
2 (two first lines of \eref{approx}) and 3 (expression \eref{approx})),
 using the cumulants \eref{cumulantsQ}. The
tails of the distribution are reproduced with increasing accuracy when 
the number of terms in the analytic expansion is increased. 
Figure \ref{fig2} displays the same for $m=11$.\\
It is to be noted that techniques similar to those used to derive 
$\langle R^n\rangle$ and $P(R)$ in section \ref{bipartite} can be
applied to derive distributions for random states drawn from
orthogonal or symplectic matrix ensembles, since the joint probability
distribution for Schmidt coefficients is of the same
form as the distribution \eref{density}.\\
\\
The author thanks CalMiP in Toulouse and Idris in Orsay for access to their 
supercomputers, and Bertrand Georgeot for reading the manuscript. 
This work was supported by the Agence Nationale de 
la Recherche (ANR project INFOSYSQQ) and the European program 
EC IST FP6-015708 EuroSQIP. 

\appendix

\section{}
\label{appA}
The aim of this appendix is to evaluate integrals of the form 
$I({\bf n})$ given by Equation \eref{inti}. The Vandermonde determinant
\eref{vandermonde} can be written
\begin{equation}
\label{vand2}
V({\bf x})
=\sum_{\sigma}\epsilon_\sigma x_{\sigma(1)}^0\ldots  x_{\sigma(p)}^{p-1},
\end{equation}
where the sum runs over all permutations on $p$ elements and 
$\epsilon_\sigma$ is the signature of the permutation $\sigma$. 
For any function $\varphi$
symmetric under permutations of the $x_i$ we have 
\begin{eqnarray}
\int_0^1 dx_1\ldots dx_p V({\bf x})^2 \varphi({\bf x})=\\
\hspace{1cm}=\nonumber\sum_{\sigma, \sigma'}\epsilon_{\sigma}\epsilon_{\sigma'}
\int_0^1 dx_1\ldots dx_p x_{\sigma(1)}^0 x_{\sigma'(1)}^0\ldots 
 x_{\sigma(p)}^{p-1}x_{\sigma'(p)}^{p-1} \varphi({\bf x})\\
\hspace{1cm}=\nonumber\sum_{\sigma, \sigma'}\epsilon_{\sigma\circ\sigma'}
\int_0^1 dx_1\ldots dx_p x_1^0 x_{\sigma\circ\sigma'(1)}^0\ldots 
 x_p^{p-1}x_{\sigma\circ\sigma'(p)}^{p-1} \varphi({\bf x})\\
\hspace{1cm}
=p!\;\int_0^1 dx_1\ldots dx_p x_1^0x_2^1\ldots x_p^{p-1}V({\bf x})
\varphi({\bf x}),\nonumber
\end{eqnarray}
with ${\bf x}=(x_1, \ldots, x_p)$. The integral \eref{inti} becomes
$I({\bf n})=\sum_\tau J(\tau({\bf n}))$ where the sum runs over all 
permutations of the $n_i$, with
\begin{equation}
\label{iequiv}
J({\bf n})=p!\int_0^1 dx_1\ldots dx_p V({\bf x})\prod_{i=1}^{p} 
x_i^{r+n_i+i-1}\delta\left(1-\sum_{i=1}^{p}x_i\right).
\end{equation}
Using the fact that
\begin{equation}
\int_0^1 dx x^a (1-x)^b=\frac{a! b!}{(a+b+1)!},
\end{equation}
a recurrence on the number of integrals shows that
\begin{equation}
\label{intmult}
\int_0^1 dx_1 \ldots dx_p x_1^{a_1}\ldots x_p^{a_p}
\delta\left(1-\sum_{i=1}^{p}x_i\right)
=\frac{a_1!a_2!\ldots a_p!}{(\sum_{i=1}^{p}a_i+p-1)!}.
\end{equation}
The Vandermonde determinant \eref{vand2} can be written as
\begin{equation}
\label{vand3}
V({\bf x})
=\sum_{\sigma}\epsilon_\sigma x^{\sigma(1)-1}_1\ldots  x^{\sigma(p)-1}_{p}.
\end{equation}
Inserting this expression in the integral \eref{iequiv} leads to a 
sum of integrals of the form \eref{intmult}, which
can be cast under
\begin{equation}
J({\bf n})=\frac{p!\prod_{i=1}^{p}(r+n_i+i-1)!}{(p^2+r p+\sum_i n_i-1)!}
\Delta({\bf n})
\end{equation}
with $\Delta({\bf n})$ a determinant defined by
\begin{equation}
\Delta({\bf n})=\left|\begin{array}{cccc}
1 & r+n_1+1 & (r+n_1+1)(r+n_1+2) & \cdots\\
1 & r+n_2+2 & (r+n_2+2)(r+n_2+3) & \cdots\\
\vdots &\vdots  &\vdots &\vdots\\
1 & r+n_p+p & (r+n_p+p)(r+n_p+p+1) & \cdots
\end{array}\right|.
\end{equation}
The determinant can be evaluated by recurrence. It finally yields
\begin{equation}
J({\bf n})=\frac{p!\prod_{i=1}^{p}(r+n_i+i-1)!}{(p^2+r p+\sum_i n_i-1)!}
\prod_{i<j}(n_j-n_i+j-i)
\end{equation}
which in turn gives $I({\bf n})$ as a sum over permutations of the $n_i$
of $J({\bf n})$.


\vspace{1cm}








\begin{thebibliography}{99} 
\bibitem{HarHayLeu}A. Harrow, P. Hayden, and D. Leung, 
Phys. Rev. Lett. {\bf 92}, 187901 (2004).

\bibitem{HayLeuShoWin}P. Hayden, D. Leung, P. Shor, and A. Winter, 
Commun. Math. Phys. {\bf 250}, 371 (2004).

\bibitem{DivLeuTer}D.~ P.~DiVincenzo, D.~W.~ Leung, and B.~M.~Terhal, 
IEEE Trans. Inf. Theory {\bf 48} (3), 580-598 (2002). 


\bibitem{WeiHel}Y. S. Weinstein and C. S. Hellberg, Phys. Rev. Lett. {\bf 95}, 
030501 (2005); Y. S. Weinstein and C. S. Hellberg, Phys. Rev. A {\bf 71}, 
014303 (2005); Y. S. Weinstein and C. S. Hellberg, Phys. Rev. A {\bf 69}, 
062301 (2004); Y. S. Weinstein and C. S. Hellberg, Phys. Rev. A {\bf 72},
 022331 (2005).


\bibitem{BanLak} J.~N.~Bandyopadhyay and A.~Lakshminarayan, 
Phys. Rev. Lett. {\bf 89}, 060402 (2002).

\bibitem{ScoCav} A.J.~Scott and C.M.~Caves, J. Phys. A: Math. Gen. {\bf 36},
9553-9576 (2003).

\bibitem{GirGeo}O. Giraud and B. Georgeot, Phys. Rev. A  {\bf 72}, 
042312 (2005).


\bibitem{BenDivSmoWoo}C.~H.~Bennett, D.~P.~DiVincenzo, J.~A.~Smolin, and 
W.~K.~Wootters, Phys. Rev. A {\bf 54}, 3824 (1996).

\bibitem{VedPleRipKni}V.~Vedral, M.~B.~Plenio, M.~A.~Rippin, and P.~L.~Knight, 
Phys. Rev. Lett. {\bf 78}, 2275 (1997).

\bibitem{VedPleJacKni}V.~Vedral, M.~B.~Plenio, K.~Jacobs, and P.~L.~Knight, 
Phys. Rev. A {\bf 56}, 4452 (1997).

\bibitem{VedPle}V.~Vedral and M.~B.~Plenio, Phys. Rev. A {\bf 57} 1619,
(1998).

\bibitem{WonChr}A.~Wong and N.~Christensen, Phys. Rev. A {\bf 63} 044301, 
(2001).

\bibitem{HilWoo}S.~Hill and W.~K.~Wootters, Phys. Rev. Lett. {\bf 78}, 5022 
(1997).

\bibitem{BenBerPopSch} C.~H.~Bennett, H.~Bernstein, S.~ Popescu, 
and B.~Schumacher, Phys. Rev. A {\bf 53}, 2046 (1996).

\bibitem{PopRoh} S.~Popescu and D.~Rohrlich, Phys. Rev. A {\bf 56}, R3319
 (1997).


\bibitem{DonHorRud} M.~J.~Donald, M.~Horodecki and O.~Rudolph , J. Math. Phys.
{\bf 43}, 4252 (2002).


\bibitem{MeyWal}A.~D.~Meyer and N.~R.~Wallach, J. Math. Phys. {\bf 43},
 4273 (2002).

\bibitem{Bre}G.~K.~Brennen, Quant. Inf. Comp. {\bf 3} 619 (2003).

\bibitem{Sco}  A.J.~Scott, Physical Review A {\bf 69}, 052330 (2004).


\bibitem{ZanZalFao} P.~Zanardi, C.~Zalka, and L.~ Faoro, Phys. Rev. A 
{\bf 62}, 030301 (2000).


\bibitem{Lub} E.~Lubkin, J. Math. Phys. (N.Y.) {\bf 19},
1028 (1978).

\bibitem{Pag}D.~N.~Page, Phys. Rev. Lett. {\bf 71}, 1291 (1993).

\bibitem{Sen}S.~Sen, Phys. Rev. Lett. {\bf 77}, 1 (1996).


\bibitem{ZycSom}K.~Zyczkowski and H.~-J.~Sommers , J. Phys. A: Math. Gen. 
{\bf 34}, 7111-7125 (2001).

\bibitem{SomZyc}H.~-J.~Sommers and K.~Zyczkowski, J. Phys. A: Math. Gen. 
{\bf 37}, 8457-8466 (2004).

\bibitem{Zyc}K.~Zyczkowski, 0606228.

\bibitem{CapSomZyc}V.~Cappellini, H.~-J.~Sommers and K.~Zyczkowski,\\ 
http://www.citebase.org/abstract?id=oai:arXiv.org:quant-ph/0605251
(2006).

\bibitem{Zni}M.~Znidaric, quant-ph/0611226.

\bibitem{MalMenLen}L.~C.~Malacarne, R.~S.~Mendes, and E.~K.~Lenzi,
Phys. Rev. E {\bf 65}, 046131 (2002).


\bibitem{Tsa}C.~Tsallis, J. Stat. Phys. {\bf 52} 479 (1988).

\bibitem{Hau}F.~Hausdorff, Math. Zeitschr. {\bf 9}, 74-109 and  280-299 (1921).

\bibitem{Gre}G.~Greaves, Numer. Math. {\bf 39}, 231-238 (1982).

\bibitem{BliMoe}S.~Blinnikov and R.~Moessner, Astron. Astrophys. Suppl. Ser. 
{\bf 130} 193-205 (1998).

\bibitem{NieChu} 
M. A. Nielsen and I. L. Chuang, Quantum computation and quantum information,  
Cambridge university press (2000).

\bibitem{Vid}G.~Vidal, J. Mod. Opt. {\bf 47} 355-376 (2000).

\bibitem{Meh}M. L. Mehta, Random Matrices (Academic Press, New York, 1991).

\bibitem{LloPag} S.~Lloyd and H. Pagels, Ann. Phys., NY {\bf 188}, 186 (1988).

\bibitem{AbrSte}M.~Abramowitz and I.~A.~Stegun, 
{\it Handbook of Mathematical Functions}, 
Dover Publications, Inc, New York (1965).

\end{thebibliography}
\end{document}